\documentclass[superscriptaddress,showpacs,twocolumn,nofootinbib]{revtex4-1}

\usepackage[normalem]{ulem}
\usepackage{graphicx}
\usepackage{color}
\usepackage{xcolor}
\usepackage{verbatim}
\usepackage{subfigure}
\usepackage{amssymb}
\usepackage{amsmath}
\usepackage{comment}
\usepackage{chngcntr}
\usepackage{appendix}
\usepackage{lipsum}
\usepackage{xr}
\usepackage{hyperref}

\usepackage{amsthm}
\usepackage{bm}
\usepackage[hang,flushmargin]{footmisc}

\theoremstyle{plain}

\begin{document}

\title{Odd Diffusivity of Chiral Random Motion}
\author{Cory Hargus}
\email{hargus@berkeley.edu}
\affiliation{Department of Chemical and Biomolecular Engineering, University of California, Berkeley, CA, USA}

\author{Jeffrey M. Epstein}
\affiliation{Department of Physics, University of California, Berkeley, CA, USA}

\author{Kranthi K. Mandadapu}
\email{kranthi@berkeley.edu}
\affiliation{Department of Chemical and Biomolecular Engineering, University of California, Berkeley, CA, USA}
\affiliation{Chemical Sciences Division, Lawrence Berkeley National Laboratory, Berkeley, CA, USA}

\begin{abstract}

Diffusive transport is characterized by a diffusivity tensor which may, in general, contain both a symmetric and an antisymmetric component. Although the latter is often neglected, we derive Green-Kubo relations showing it to be a general characteristic of random motion breaking time-reversal and parity symmetries, as encountered in chiral active matter. In analogy with the odd viscosity appearing in chiral active fluids, we term this component the odd diffusivity. We show how odd diffusivity emerges in a chiral random walk model, and demonstrate the applicability of the Green-Kubo relations through molecular dynamics simulations of a passive tracer particle diffusing in a chiral active bath.

\vspace{5mm}
\noindent DOI: \href{https://doi.org/10.1103/PhysRevLett.127.178001}{10.1103/PhysRevLett.127.178001}
\end{abstract}

\maketitle

\vspace{0.1in}
\noindent\textbf{\textit{Introduction.}}
Among the historic successes of nonequilibrium statistical mechanics is the explanation of macroscopic transport phenomena in terms of microscopic fluctuations occurring at equilibrium~\cite{Onsager1931a,Onsager1931b,Prigogine,deGroot1951,Groot1984}.
More recent efforts aim to generalize this framework to include systems whose steady states are not Boltzmann distributed, and whose dynamics are not determined by Hamiltonian-conserving forces.
A major impetus for this generalization is the study of active matter, \textit{i.e.} systems composed of particles that are propelled by microscopic driving forces and thus maintained out of equilibrium.

Chiral active matter is composed of particles driven by microscopic torques and may be synthetic, as in the case of active colloids~\cite{Kummel2013,Nourhani2016,Soni2019,Witten2020}, or biological, as in the case of certain bacteria, algae, and spermatozoa~\cite{Diluzio2005,Drescher2009,Riedel2005}.
Such systems have been shown to exhibit emergent transport behavior reminiscent of their equilibrium counterparts, yet with striking differences.
For instance, chiral active fluids may exhibit Newtonian constitutive behavior, but with a novel viscosity coefficient termed the odd (or Hall) viscosity emerging as a consequence of breaking time-reversal and parity symmetries at the level of stress fluctuations~\cite{banerjee2017odd,Epstein2020,Hargus2020,han2020statistical}.
In this Letter we examine an analogous quantity appearing in the context of diffusive transport.

In dilute solutions, Fick's law posits the linear constitutive relation
\begin{equation}\label{eq:ficks-law}
    \bm{J} = -\mathbf{D} \cdot \bm{\nabla} C
\end{equation}
between the diffusive flux $\bm{J}$ and the concentration gradient $\bm{\nabla} C$,
with $\mathbf{D}$ being a rank-two diffusivity tensor.
In general $\mathbf{D}$ may contain both a symmetric and antisymmetric part.
We term the latter the ``odd diffusivity,'' emphasizing its connection to odd viscosity.
Just as odd viscosity generates normal stresses perpendicular to shear flow, odd diffusivity generates fluxes perpendicular to concentration gradients.
Like odd viscosity~\cite{avron1995viscosity,Avr98,banerjee2017odd,Epstein2020,Hargus2020,han2020statistical}, we will show odd diffusivity to emerge as a consequence of breaking time-reversal and parity symmetries at the level of microscopic fluctuations.

\begin{figure}[!t]
    \centering
    \includegraphics[width=.5\textwidth]{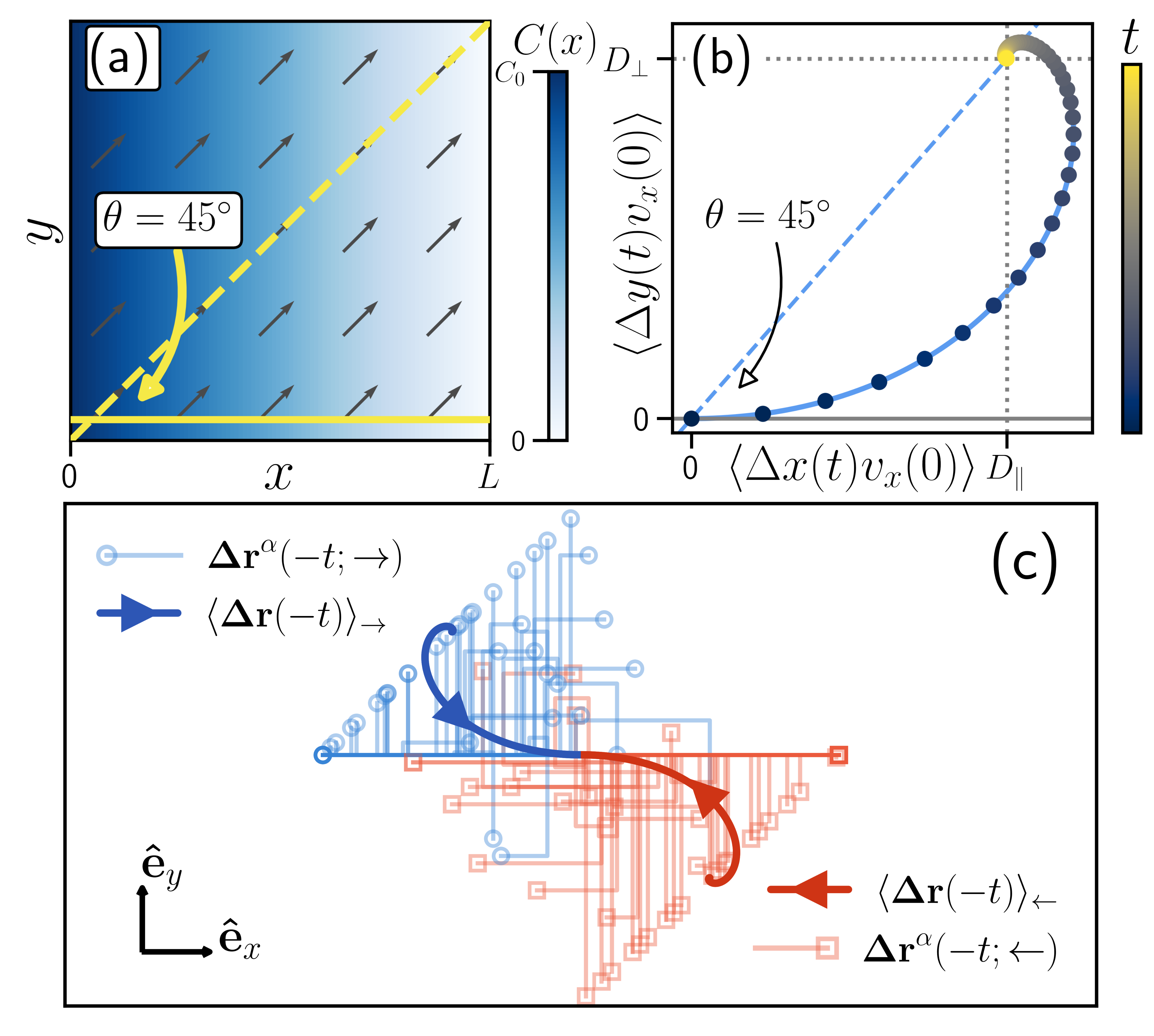}
    \caption{
    Relationship between odd diffusivity and chirality of particle trajectories in a left-turning random walk ($\Gamma_1 = 1, \Gamma_2=\Gamma_3=0$).
    (a) A linear concentration gradient induces a uniform flux field (arrows) with a perpendicular component due to $D_\perp$.
    (b) Logarithmic spiral form of the position-velocity correlation functions from equations~\eqref{eq:CRW-x-average}-\eqref{eq:CRW-y-average}. The Green-Kubo relations~\eqref{eq:gk-1}-\eqref{eq:gk-2} specify that the $x$- and $y$-coordinates converge to the two diffusivity coefficients as $t \rightarrow \infty$, while the angle $\theta$ is identical to that in (a), as annotated.
    (c) Random sample of 50 time-reversed trajectories $\bm{\Delta r}^\alpha(-t)$ satisfying either $\bm{v}^\alpha(0)=v_0\hat{\bm{e}}_x$ (indicated by $\rightarrow$) or $\bm{v}^\alpha(0)=-v_0\hat{\bm{e}}_x$ (indicated by $\leftarrow$) for $t \in [0,\Gamma_1^{-1}]$ together with the subensemble-averaged trajectories $\langle \bm{\Delta r}^\alpha(-t) \rangle_\rightarrow$ and $\langle \bm{\Delta r}^\alpha(-t) \rangle_\leftarrow$ for $t \in [0,\infty)$.
    }
    \label{fig:subensembles}
\end{figure}

For simplicity, we examine odd diffusivity in isotropic systems.
As there exists no rank-two tensor in three dimensions which is both isotropic and antisymmetric~\cite{Epstein2020} we restrict our attention to two-dimensional diffusion, where the diffusivity tensor takes the form
\begin{equation} \label{eq:diffusion-tensor}
    D_{ij} = D_\parallel \delta_{ij} - D_\perp \epsilon_{ij}
      = \begin{bmatrix}
D_\parallel & -D_\perp\\
D_\perp & D_\parallel
\end{bmatrix}
\,.
\end{equation}
Here, $\delta_{ij}=\delta_{ji}$ is the symmetric Kronecker delta and $\epsilon_{ij} = -\epsilon_{ji}$ is the antisymmetric Levi-Civita permutation tensor.
$D_\parallel$ is the ordinary isotropic diffusivity coefficient driving flux from regions of high to low concentration while $D_\perp$ is the odd diffusivity driving flux in the perpendicular direction (as in Figure~\ref{fig:subensembles}a).
Combining~\eqref{eq:ficks-law} and~\eqref{eq:diffusion-tensor} with the continuity equation $\partial_t C = -\bm{\nabla} \cdot \bm{J}$ yields the diffusion equation
\begin{equation}\label{eq:diffusion-equation}
    \partial_t C = D_\parallel \nabla^2 C\,,
\end{equation}
which is unaffected by the divergence-free fluxes produced by $D_\perp$.
Thus, while $D_\perp$ may influence $C$ in the presence of boundary conditions involving fluxes (\textit{e.g.}\ impermeable obstacles, see Appendix A.1), $D_\perp$ cannot affect $C$ for boundary conditions involving solely the concentration.

Past studies of odd diffusivity have generally been limited to equilibrium systems, most commonly systems of charged particles in magnetic fields.
Such systems acquire an antisymmetric component of both the diffusivity tensor and the mobility tensor, which describes the current response to an electric field.
This is the basis of the Hall effect, and has consequences for the transport of confined plasmas and cosmic rays~\cite{Townsend1912,Landauer1953,Spitzer1956,Bieber1997,Giacalone1999,Abdoli2020,Bonella2017,Coretti2018}.
Odd diffusivity has also been recognized in certain mathematical models of chiral random walks~\cite{Larralde1997,Hijikata2015}, and in convection-diffusion processes in chiral porous media~\cite{Koch1987}.

In this Letter we suggest a unifying framework within which to understand these phenomena, which extends beyond equilibrium.
We begin by asking: given that the existence of odd diffusivity is compatible with the macroscopic theory of diffusion, what microscopic conditions are necessary for it to appear?
Through deriving a Green-Kubo relation for the odd diffusivity, we will show that it emerges in systems breaking time-reversal and parity symmetries,
as characterized by chiral random motion of particle trajectories.
Odd diffusivity is thus characteristic of a broad range of diffusive processes, and of particular interest for out-of-equilibrium systems such as chiral active matter, where time-reversal symmetry can be broken by microscopic driving forces.
We validate the derived Green-Kubo relations exactly for a model chiral random walk and numerically in active matter simulations, demonstrating good agreement with direct measurements of the flux in response to an imposed concentration gradient.

\vspace{0.1in}
\noindent\textbf{\textit{Green-Kubo relations.}}
We now proceed to obtain Green-Kubo relations for $D_{ij}$.
We follow an approach similar in spirit to the celebrated work of Einstein, Smoluchowski and others~\cite{Einstein1905,Smoluchowski1906}, which connected molecular-scale Brownian motion with the macroscopic diffusion equation~\eqref{eq:diffusion-equation}, and we will rely on similar arguments about the separation of timescales.
However, because the odd diffusivity $D_\perp$ does not contribute to equation~\eqref{eq:diffusion-equation}, such an approach can yield no information about $D_\perp$.
The same is true when taking as a starting point the Onsager regression hypothesis~\cite{Onsager1931a,Onsager1931b,Kubo1957b}, itself formulated upon equation~\eqref{eq:diffusion-equation}, as in a recent derivation of Green-Kubo relations for the odd viscosity~\cite{Epstein2020}.
Accordingly, rather than considering the time evolution of the concentration \textit{via} the diffusion equation~\eqref{eq:diffusion-equation}, we will instead directly examine the microscopic basis of the fluxes appearing in the constitutive law~\eqref{eq:ficks-law}, similar to the route taken in linear response theory~\cite{EvansMorris}.
In doing so, however, we will not require any linear response relation between the diffusivity and the mobility.

We begin by considering a dilute solution of particles undergoing random motion, \textit{e.g.}\ due to collisions with a solvent bath.
Let $f(\bm{r}, \bm{v}, t)$ indicate the probability density of finding a particle at position $\bm{r}$ with velocity $\bm{v}$ at time $t$.
The local, instantaneous flux $\bm{J}(\bm{r},t)$ is then defined as
\begin{equation} \label{eq:local-instantaneous-flux}
    \bm{J}(\bm{r},t) = \int d\bm{v}\ f(\bm{r}, \bm{v}, t) \bm{v} \,.
\end{equation}

Let us now consider the subensemble of all single-particle trajectories compatible with the conditions $\bm{r}^\alpha(t) = \bm{r}$ and $\bm{v}^\alpha(t) = \bm{v}$, where $\alpha$ is an index over trajectories.
As particles cannot be created or destroyed, continuity requires that
\begin{equation} \label{eq:trajectory-continuity}
    f(\bm{r}, \bm{v}, t) = \big\langle f\big(\bm{r}^\alpha(t-\tau), \bm{v}^\alpha(t-\tau), t - \tau \big) \big\rangle_{\substack{\bm{r}^\alpha(t) = \bm{r} \\ \bm{v}^\alpha(t) = \bm{v}}}\,,
\end{equation}
where $\big\langle \cdot \big\rangle_{\substack{\bm{r}^\alpha(t) = \bm{r} \\ \bm{v}^\alpha(t) = \bm{v}}}$ denotes an average over all trajectories leading into point $\bm{r}$ with velocity $\bm{v}$ at time $t$.
Suppose there exists a correlation timescale $\tau_c$, such that for $\tau \gg \tau_c$ a particle's velocity $\bm{v}^\alpha(t)$ is uncorrelated with its earlier value $\bm{v}^\alpha(t - \tau)$ and thus becomes distributed according to the unconditional probability density function $\phi(\bm{v})$, which we assume to be independent of $t$ (stationary) and $\bm{r}$ (translationally invariant).
Then, for $\tau \gg \tau_c$, equation~\eqref{eq:trajectory-continuity} factorizes to
\begin{equation} \label{eq:trajectory-continuity-factorized}
    f(\bm{r}, \bm{v}, t) = \phi(\bm{v}) \langle C\big(\bm{r}^\alpha(t-\tau), t - \tau \big) \rangle_{\substack{\bm{r}^\alpha(t) = \bm{r} \\ \bm{v}^\alpha(t) = \bm{v}}}\,,
\end{equation}
where the concentration $C(\bm{r}, t) = \int d\bm{v}\ f(\bm{r}, \bm{v}, t)$.

Let the timescale over which the system relaxes from a state of nonuniform concentration be denoted $\tau_r$, \textit{e.g.} ${\tau_r \approx L^2 / D_\parallel}$, for the macroscopic length $L$ describing the variation in $C(\bm{r}, t)$.
We now assume that $\tau$ may be chosen to satisfy the separation of timescales
\begin{equation}
    \tau_c \ll \tau \ll \tau_r \,,
 \end{equation}
following Einstein, Smoluchowski, Kubo and others~\cite{Einstein1905,Smoluchowski1906,Kubo1957,Kubo1957b,Epstein2020}.
With these assumptions, the subensemble-averaged concentration appearing in equation~\eqref{eq:trajectory-continuity-factorized} may be approximated by expanding about $\bm{r}$ to first order and about $t$ to zeroth order
\begin{align}\label{eq:concentration-expansion}
    \begin{split}
    &\bigg\langle C\big(\bm{r}^\alpha(t-\tau), t-\tau \big) \bigg\rangle_{\substack{\bm{r}^\alpha(t) = \bm{r} \\ \bm{v}^\alpha(t) = \bm{v}}} \\
     &\approx C(\bm{r}, t) + \big\langle \bm{r}^\alpha(t - \tau) - \bm{r}^\alpha(t) \big\rangle_{\substack{\bm{r}^\alpha(t) = \bm{r} \\ \bm{v}^\alpha(t) = \bm{v}}} \cdot \bm{\nabla} C(\bm{r}, t) \,.
    \end{split}
\end{align}

Noting the relationship between a particle's displacement and its velocity
\begin{equation}\label{eq:displacement-velocity-relation}
    \bm{r}^\alpha(t - \tau) - \bm{r}^\alpha(t) = - \int_0^\tau dt'\ \bm{v}^\alpha(t - t')
\end{equation}
and inserting the results of equations~\eqref{eq:trajectory-continuity-factorized}-\eqref{eq:displacement-velocity-relation} into equation~\eqref{eq:local-instantaneous-flux} yields
\begin{align}\label{eq:flux-conditional-2}  
        &\bm{J}(\bm{r}, t) = \int d\bm{v}\ \phi(\bm{v}) \bm{v}\ \times \notag \\
        & \hspace{5mm} \bigg[ C(\bm{r}, t) - \int_0^{\tau} dt'\ \big\langle \bm{v}^\alpha(t - t') \big\rangle_{\substack{\bm{r}^\alpha(t) = \bm{r} \\ \bm{v}^\alpha(t) = \bm{v}}} \cdot \bm{\nabla} C(\bm{r}, t) \bigg]\notag \\
        &= -\int_0^{\tau} dt'\ \langle \bm{v}(t) \otimes \bm{v}(t - t') \rangle \cdot \bm{\nabla} C(\bm{r}, t) \,,
\end{align}
with $\otimes$ indicating the dyadic product.
The convective term proportional to $C(\bm{r}, t)$ vanishes under the assumption that $\phi(\bm{v})$ is unbiased, \textit{i.e.} $\int d\bm{v}\, \phi(\bm{v}) \bm{v}=0$.
The second equality in~\eqref{eq:flux-conditional-2} follows from the definition of the conditional expectation.
The condition $\bm{r}^\alpha(t) = \bm{r}$ has been dropped due to the assumption of translational invariance;
consequently, the average in the final expression is taken over all trajectories.
Comparing with the constitutive relation~\eqref{eq:ficks-law}, we conclude
\begin{equation}
    D_{ij} = \int_0^{\tau} dt'\ \langle v_i(t)  v_j(t - t') \rangle \,.
\end{equation}
Invoking stationarity to set $\langle v_i(t)  v_j(t - t') \rangle = \langle v_i(t')  v_j(0) \rangle$
and carrying out the limit $\tau \rightarrow \infty$ due to the requirement $\tau \gg \tau_c$ yields the Green-Kubo relations
\begin{equation}\label{eq:gk-full-tensor}
    D_{ij} = \int_0^{\infty} dt\ \langle v_i(t)  v_j(0) \rangle\,.
\end{equation}
These relations hold independently for each component of the diffusivity tensor, including any antisymmetric part.
Considering the specific form of $D_{ij}$ in~\eqref{eq:diffusion-tensor}, we may contract with $\delta_{ij}$ and $\epsilon_{ij}$ to obtain
\begin{align}\label{eq:gk-1}
    \begin{split}
    2D_\parallel &= \int_0^{\infty} dt\, \langle v_i(t)  v_j(0) \rangle \delta_{ij} \\
    &=\lim_{t \rightarrow \infty} \langle \Delta r_i(t)  v_j(0) \rangle \delta_{ij}
    = \lim_{t \rightarrow \infty} \frac{1}{2t} \langle | \bm{\Delta r}(t) |^2\rangle \,,
    \end{split} \\
     \label{eq:gk-2}
    \begin{split}
    2D_\perp &= -\int_0^{\infty} dt\, \langle v_i(t)  v_j(0) \rangle \epsilon_{ij} \\
    &=-\lim_{t \rightarrow \infty} \langle \Delta r_i(t)  v_j(0) \rangle \epsilon_{ij}\,.
    \end{split}
\end{align}

The first equality in equations~\eqref{eq:gk-1} and~\eqref{eq:gk-2} is of the usual Green-Kubo form~\cite{Kubo1957,Kubo1957b}. In the second equality the integral has been carried out, permitting a geometric interpretation of the two diffusion coefficients in terms of the position-velocity correlation functions (as in Figure~\ref{fig:subensembles}b).
The third equality in~\eqref{eq:gk-1} is the well-known relationship between $D_\parallel$ and the mean squared displacement; note that no such relation exists for $D_\perp$ due to its absence from the diffusion equation~\eqref{eq:diffusion-equation}.

The antisymmetric tensor $\epsilon_{ij}$ in equation~\eqref{eq:gk-2} projects out the time-reversal-symmetric and even-parity part of the correlation function, indicating that whereas $D_\parallel$ is even under time reversal and parity inversion, $D_\perp$ is odd under both operations.
Onsager's reciprocal relations~\cite{Onsager1931a,Onsager1931b} similarly require that transport coefficient tensors be symmetric as a consequence of time-reversal symmetry.
It should be noted however that $D_\perp$, being non-dissipative, is not compatible with entropic arguments pertaining to the reciprocal relations, an issue that was previously discussed in a Fokker-Planck context~\cite{Tomita1974,Tomita1974Erratum}.
The Green-Kubo relation \eqref{eq:gk-2} provides, instead, a direct statement of how time-reversal symmetry should be broken for odd diffusivity to appear.

In equilibrium systems, the diffusivity and the mobility are connected by the Einstein relation.
In such systems, the Green-Kubo relation~\eqref{eq:gk-2} may be shown from linear response theory~\cite{EvansMorris}.
The derivation above shows that equation~\eqref{eq:gk-2} can be applied even to inherently nonequilibrium systems such as active matter,
where effective Einstein relations may exist under special circumstances~\cite{Baiesi2009,Baiesi2013,Maes2011,DalCengio2019,Shakerpoor2021}, but in general need not.
Consequently, odd diffusivity can arise even in cases where the antisymmetric mobility vanishes (as demonstrated in Appendix A.3 for a chiral active Brownian particle), or where mobility has no physical meaning, as in cases of animal navigation with a documented steering bias~\cite{Komin2004,Codling2008,Souman2009,Bestaven2012,romanczuk2012active}.

\vspace{0.1in}
\noindent\textbf{\textit{Chiral random walk.}}
To illustrate the microscopic origins of $D_\perp$ and $D_\parallel$, consider a particle which moves at a constant speed $v_0$ and reorients by turning left, reversing direction, or turning right at random intervals with frequency $\Gamma_1$, $\Gamma_2$ and $\Gamma_3$, respectively.
Between these changes in direction, the particle moves in a straight line.

We may understand the diffusive behavior of this model by decomposing the probability density $P(x,y,t)$ of the particle sitting at coordinates $(x,y)$ at time $t$ into a sum of joint probabilities associated with the four possible directions of motion: $P(x,y,t) = P_\rightarrow(x,y,t) + P_\uparrow(x,y,t) + P_\leftarrow(x,y,t) + P_\downarrow(x,y,t)$.
By considering the continuity of these joint probabilities, we arrive at the coupled master equations~\cite{Risken1989}
\begin{align}
    \label{eq:master-1}
    \partial_t P_\rightarrow &= \Gamma_1 P_\downarrow + \Gamma_2 P_\leftarrow + \Gamma_3 P_\uparrow - \gamma P_\rightarrow -v_0 \partial_x P_\rightarrow \,, \\
    \label{eq:master-2}
    \partial_t P_\uparrow &= \Gamma_1 P_\rightarrow + \Gamma_2 P_\downarrow + \Gamma_3 P_\leftarrow - \gamma P_\uparrow -v_0 \partial_y P_\uparrow \,, \\
    \label{eq:master-3}
    \partial_t P_\leftarrow &= \Gamma_1 P_\uparrow + \Gamma_2 P_\rightarrow + \Gamma_3 P_\downarrow - \gamma P_\leftarrow + v_0 \partial_x P_\leftarrow \,, \\
    \label{eq:master-4}
    \partial_t P_\downarrow &= \Gamma_1 P_\leftarrow + \Gamma_2 P_\uparrow + \Gamma_3 P_\rightarrow - \gamma P_\downarrow + v_0 \partial_y P_\downarrow \,,
\end{align}
where $\gamma = \Gamma_1 + \Gamma_2 + \Gamma_3$.
Suppose we are interested in a steady state in which concentration varies only in the $x$-direction. Then, from equation~\eqref{eq:local-instantaneous-flux}, we may define
\begin{align}
    J_x(x) &=  v_0 \langle P_\rightarrow(x) - P_\leftarrow(x) \rangle\,, \\
    J_y(x) &=  v_0 \langle P_\uparrow(x) - P_\downarrow(x) \rangle \,,
\end{align}
and, upon subtracting equation~\eqref{eq:master-4} from~\eqref{eq:master-2} and averaging, obtain
\begin{equation}
    \partial_t J_y(x) = 0 = (\Gamma_1 - \Gamma_3) J_x(x) - (\gamma + \Gamma_2) J_y(x)\,.
\end{equation}
Solving for the ratio $J_y(x) / J_x(x)$, we find
\begin{equation}\label{eq:alpha-nemd}
\frac{J_y(x)}{J_x(x) } = \frac{D_\perp}{D_\parallel} = \frac{\Gamma_1 - \Gamma_3}{\gamma + \Gamma_2}\,.
\end{equation}
Examining this expression we note that $D_\perp \ne 0$ whenever $\Gamma_1 \ne \Gamma_3$, indicating a preference between left and right turns, \textit{i.e.}\ chirality of random motion.

We now consider the Green-Kubo relation~\eqref{eq:gk-1} for this model.
Recognizing that only four velocity states are possible, we expand the correlation functions as
\begin{align}\label{eq:chiral-rw-gk-1}
    D_\parallel &= \lim_{t \rightarrow \infty} \frac{1}{2} \langle \Delta r_i(t) v_j(0) \rangle \delta_{ij}\notag \\
    = &\lim_{t \rightarrow \infty} \frac{1}{8} v_0 \big[ \langle x(t) \rangle_\rightarrow  - \langle x(t) \rangle_\leftarrow + \langle y(t) \rangle_\uparrow - \langle y(t) \rangle_\downarrow \big] \notag \\
    = &\lim_{t \rightarrow \infty} \frac{1}{2} v_0 \langle x(t) \rangle_\rightarrow \,,
\end{align}
where $\langle \cdot \rangle_\rightarrow$ indicates an average conditioned on the particle initially moving to the right from the origin.
The other terms $\langle \cdot \rangle_\uparrow$, $\langle \cdot \rangle_\leftarrow$ and $\langle \cdot \rangle_\downarrow$ follow the same notational convention.
The simplification on the final line is due to isotropy.
Likewise, from equation~\eqref{eq:gk-2},
\begin{equation}\label{eq:chiral-rw-gk-2}
    D_\perp = \lim_{t \rightarrow \infty} \frac{1}{2} v_0 \langle y(t) \rangle_\rightarrow \,.
\end{equation}
The averages are obtained by solving equations~\eqref{eq:master-1} through~\eqref{eq:master-4} with the initial condition $P_\rightarrow(x,y,0) = \delta(x)\delta(y)$ (see Appendix A.1). In doing so, we find that the mean trajectory is a logarithmic spiral, \textit{i.e.}
\begin{align}
    \label{eq:CRW-x-average}
    \langle x(t) \rangle_\rightarrow &= v_0 \frac{\nu - e^{-\nu t} \big( \nu \cos(\omega t) + \omega \sin(\omega t) \big)}{\nu^2 + \omega^2}\\
    \label{eq:CRW-y-average}
    \langle y(t) \rangle_\rightarrow &= v_0 \frac{\omega - e^{-\nu t} \big( \omega \cos(\omega t) - \nu \sin(\omega t) \big)}{\nu^2 + \omega^2}
\end{align}
where for compactness we have defined $\omega = \Gamma_1 - \Gamma_3$ and $\nu = \Gamma_1 + 2\Gamma_2 + \Gamma_3$.
This logarithmic spiral functional form, shown in Figure~\ref{fig:subensembles}b,
is remarkably common, appearing in the mean trajectories of charged particles diffusing in a magnetic field~\cite{Townsend1912,Spitzer1956,Abdoli2020a,Vuijk2020}, as well as those of chiral active colloids~\cite{Nourhani2016,Kummel2013} and certain biological systems~\cite{VanTeeffelen2008,Codling2008}. Inserting equations~\eqref{eq:CRW-x-average}-\eqref{eq:CRW-y-average} into~\eqref{eq:chiral-rw-gk-1}-\eqref{eq:chiral-rw-gk-2} yields
\begin{align}
\label{eq:Dpar-crw}
    2D_\parallel &= v_0^2 \frac{\nu}{\nu^2 + \omega^2}\,, \\
\label{eq:Dperp-crw}
    2D_\perp &= v_0^2 \frac{\omega}{\nu^2 + \omega^2}\,,
\end{align}
in agreement with equation~\eqref{eq:alpha-nemd}, showing the emergence of $D_\perp$ when chirality is present ($\omega \ne 0$).

Figure~\ref{fig:subensembles} illustrates the origins of odd diffusivity in a chiral random walk which permits only left turns ($\Gamma_1=1, \Gamma_2=\Gamma_3=0$), for which $D_\parallel = D_\perp$, from equations~\eqref{eq:Dpar-crw}-\eqref{eq:Dperp-crw}.
Figure~\ref{fig:subensembles}a displays the steady-state solution to equations~\eqref{eq:ficks-law}-\eqref{eq:diffusion-equation} for diffusion between two reservoirs with concentrations $C(x\text{=}0) = C_0$ and $C(x\text{=}L) = 0$, resulting in a linear concentration profile $C(x) = C_0(1 - x/L)$ and uniform flux $\bm{J} = \frac{C_0}{L}\big[ D_\parallel \hat{\bm{e}}_x + D_\perp \hat{\bm{e}}_y \big]$ with a nonzero $y$-component due to odd diffusivity.
In the presence of impermeable boundaries this solution must be modified, with $D_\perp$ affecting not only the flux but also the concentration, as shown in Appendix A.1.
Figure~\ref{fig:subensembles}b plots the position-velocity correlation functions entering into the Green-Kubo relations~\eqref{eq:gk-1} and~\eqref{eq:gk-2}.
Finally, Figure~\ref{fig:subensembles}c shows a random sample from the subensembles of time-reversed trajectories $\bm{\Delta r}^\alpha(-t)$ passing through the origin at time $t=0$ with either $\bm{v}^\alpha(0) = +v_0\hat{\bm{e}}_x$ or $\bm{v}^\alpha(0) = -v_0\hat{\bm{e}}_x$.
Due to chirality, the paths in these two subensembles lead backwards in time to regions differing not only in the $x$- but also the $y$-coordinate, so that a gradient in the $y$-direction generates a flux in the $x$-direction.
This is the microscopic basis of odd diffusivity.

\vspace{0.1in}
\noindent\textbf{\textit{Diffusion in a chiral active bath.}}
\begin{figure}[!t]
    \centering
    \includegraphics[width=.48\textwidth]{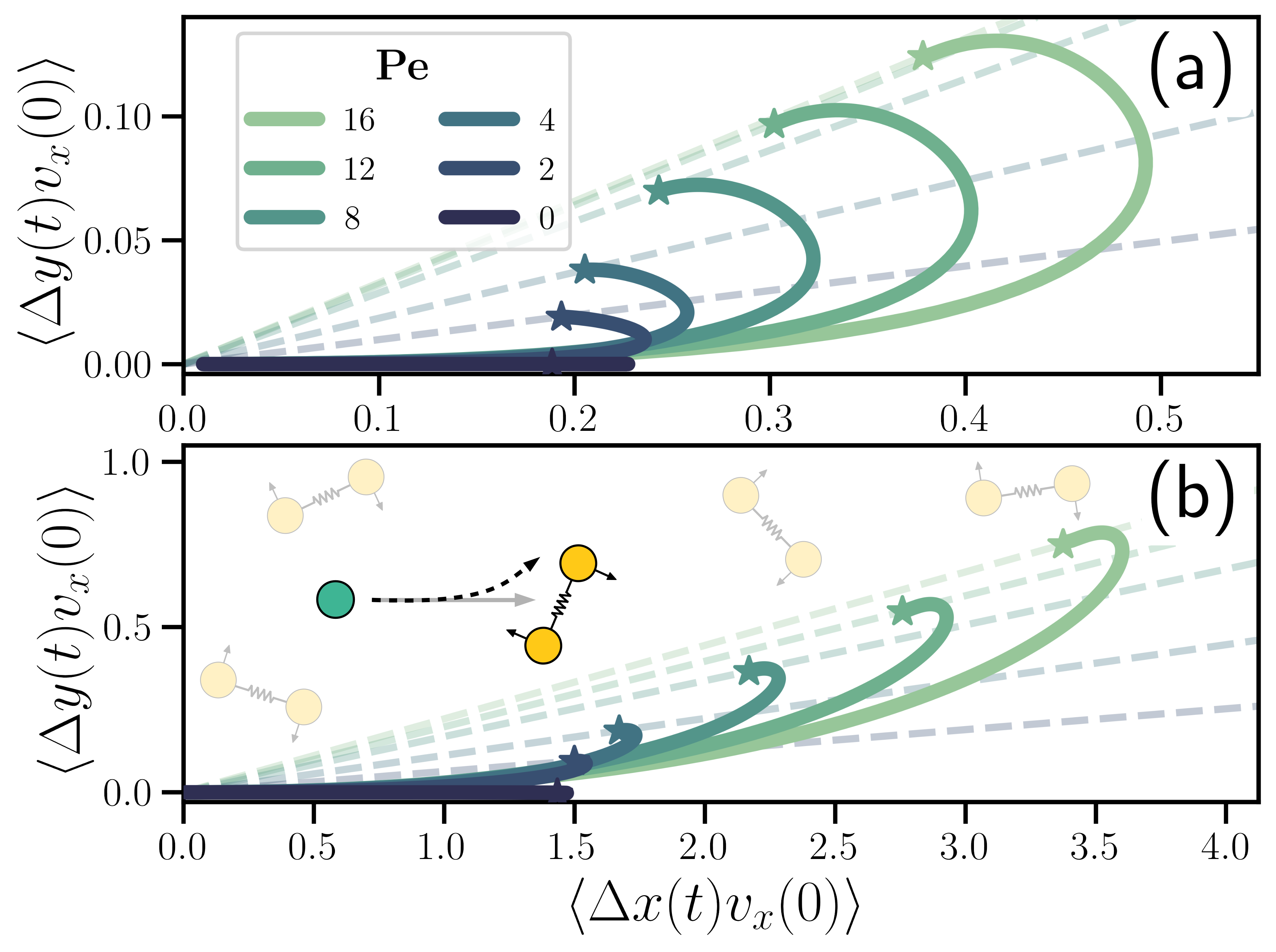}
    \caption{Position-velocity correlation functions computed from molecular dynamics simulations of a passive tracer in a chiral active dumbbell bath with density $\rho_\mathrm{bath}=0.4$ (a) and $\rho_\mathrm{bath}=0.1$ (b).
    Stars mark converged values as $t \rightarrow \infty$.
    Both $D_\perp$ and $D_\parallel$ increase with $\mathrm{Pe}$, as does the ratio $D_\perp/D_\parallel$, as indicated by dashed lines.
    The inset in (b) depicts the model system.
    }
    \label{fig:dumbbell_bath_spirals}
\end{figure}
Several recent studies have described novel behavior of the symmetric diffusivity $D_\parallel$~\cite{Weber2011,Volpe2014,Sevilla2016,Kanazawa2020} as well as an antisymmetric mobility~\cite{Nourhani2013,Kogan2016,Reichhardt2019,Hosaka2021} in active systems.
In this section, we study the odd diffusivity of a passive tracer particle dissolved in a two-dimensional chiral active fluid composed of torqued dumbbells, which was found in previous studies to exhibit odd viscosity and an asymmetric hydrostatic stress~\cite{klymko2017statistical,Hargus2020}.
The positions $\bm{r}_i$ and velocities $\bm{v}_i$ of particle $i$ evolve according to underdamped Langevin dynamics
\begin{align} \label{eq:langevin-dynamics}
    \begin{split}
        \dot{\bm{r}}_i &= \bm{v}_i\,, \\
        \dot{\bm{v}}_i &= -\frac{\partial}{\partial \bm{r}_i}U + \bm{f}^A_i - \zeta \bm{v}_i + \bm{\eta}_i \,,
    \end{split}
\end{align}
with particle masses set to one.
Here, $-\frac{\partial}{\partial \bm{r}_i}U$ is the conservative force on particle $i$ due to interactions
(see Appendix A.2 for model and simulation details).
$\bm{f}^A_i$ is a nonconservative active force inducing rotation of the dumbbell.
$\zeta$ is the dissipative bath friction and $\bm{\eta}_i$ are the bath fluctuations, modeled as Gaussian white noise characterized by $\langle \bm{\eta}_i \rangle = 0$ and $\langle \bm{\eta}_i(t) \otimes \bm{\eta}_j(t') \rangle = 2 k_\mathrm{B} T \zeta \delta_{ij} \delta(t-t') \mathbf{I}$, where $k_\mathrm{B} T$ is the bath temperature and $\mathbf{I}$ is the identity matrix.
In all simulations the density of active dumbbells, $\rho_\mathrm{bath}$, is spatially homogeneous.
The magnitude of $f^A = |\bm{f}^A_i|$ relative to thermal fluctuations is quantified by a non-dimensional P\'eclet number defined as
$\mathrm{Pe} = \frac{2 f^A d}{k_\mathrm{B} T}$,
where $d$ is the equilibrium dumbbell bond length.

\begin{figure}[!t]
    \centering
    \includegraphics[width=.48\textwidth]{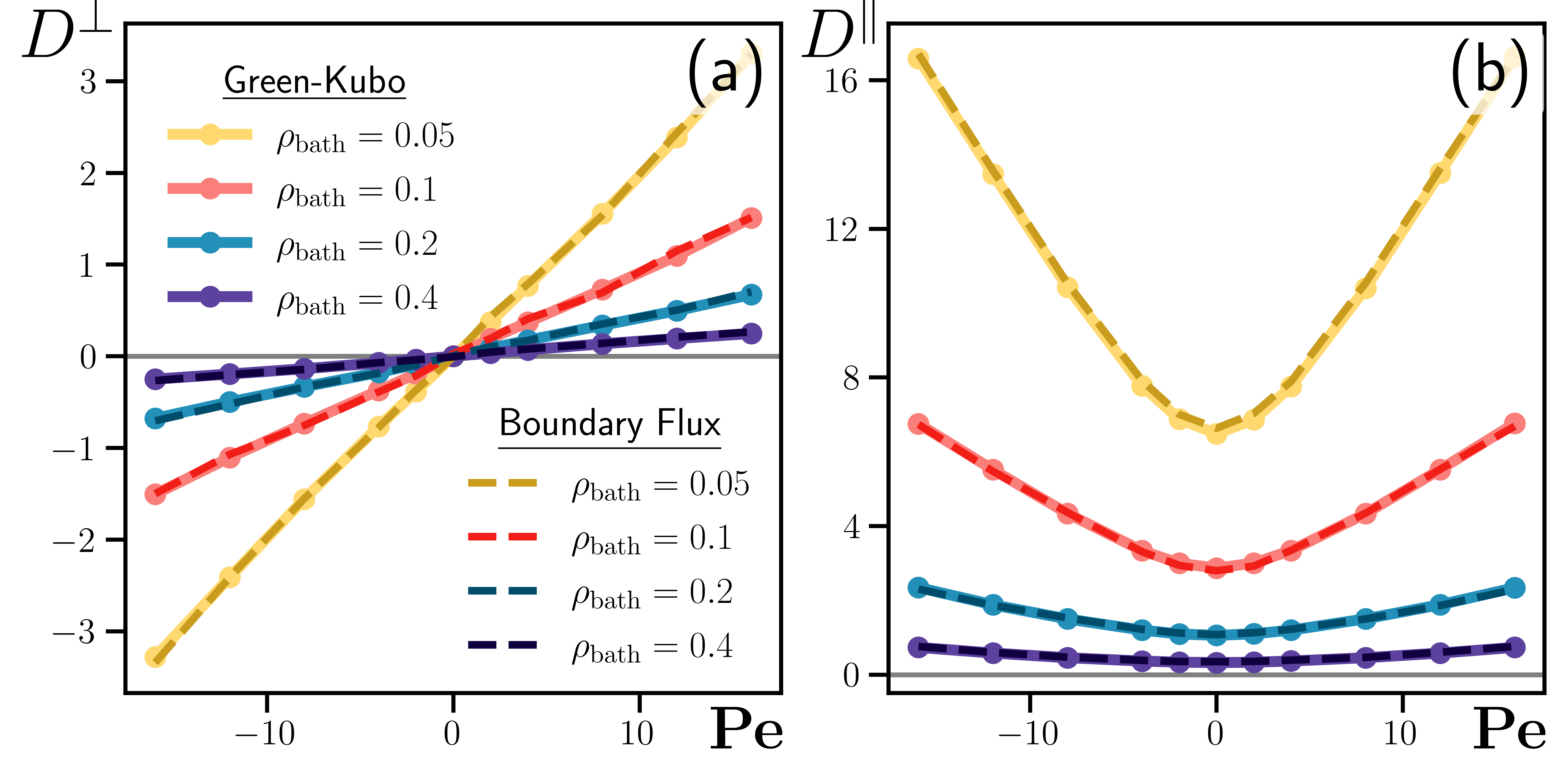}
    \caption{Comparison of the diffusion coefficients $D_\perp$ (a) and $D_\parallel$ (b) computed from the Green-Kubo relations (solid lines) with those measured in boundary-driven flux simulations (dashed lines) for several densities of the active dumbbell bath $\rho_\mathrm{bath}$ and values of $\mathrm{Pe}$. Error bars are smaller than the symbols.
    }
    \label{fig:gk_vs_nemd}
\end{figure}

Molecular dynamics simulations~\cite{lammps,Weeks71} with fully periodic boundaries allow for the measurement of the position-velocity correlation functions, which are plotted in Figure~\ref{fig:dumbbell_bath_spirals}.
We have taken the convention that $\mathrm{Pe} > 0$ corresponds to clockwise rotation of the dumbbells, which induces counterclockwise motion of the passive tracer, as depicted in the inset of Figure~\ref{fig:dumbbell_bath_spirals}b. When $\mathrm{Pe} \ne 0$, an antisymmetric part of the correlation function appears, with a shape resembling the logarithmic spirals identified in the chiral random walk model (Figure~\ref{fig:subensembles}b) and magnitude depending strongly on the density of the active dumbbell bath. The resulting Green-Kubo estimates of $D_\perp$ and $D_\parallel$ are plotted in Figures~\ref{fig:gk_vs_nemd}a and~\ref{fig:gk_vs_nemd}b for a range of active bath densities, where $D_\perp$ is seen to be an odd function of $\mathrm{Pe}$ while $D_\parallel$ is an even function of $\mathrm{Pe}$.

To validate the Green-Kubo relations, we independently performed boundary-driven flux simulations in which passive tracer particles at high dilution were introduced at the left boundary of the simulation box and removed from the right boundary at a constant rate, while the top and bottom boundaries remained periodic.
The resulting steady state exhibits a uniform concentration gradient in the $x$-direction, and uniform flux with a $y$-component emerging for $\mathrm{Pe} \ne 0$ (see Appendix A.2).
The diffusion coefficients $D_\perp$ and $D_\parallel$ were then computed directly from the constitutive relations~\eqref{eq:ficks-law} and~\eqref{eq:diffusion-tensor}.
The resulting values are plotted in Figure~\ref{fig:gk_vs_nemd} against the Green-Kubo predictions, demonstrating good agreement.
We note that this system exhibits an antisymmetric part of the mobility, but with no apparent Einstein relation connecting this quantity to the odd diffusivity (see Appendix A.3).

\vspace{0.1in}
\noindent\textbf{\textit{Conclusion.}}
Ordinarily, isotropic diffusion involves fluxes parallel to concentration gradients.
In general, however, there may emerge fluxes in the perpendicular direction.
This behavior appears as an antisymmetric part of the diffusivity tensor, which we have termed odd diffusivity.
From a first-principles consideration of the microscopic basis of the constitutive relations describing these perpendicular fluxes, we have derived a Green-Kubo relation for odd diffusivity, showing it to exist only when time-reversal and parity symmetries are broken, whether in or out of equilibrium.
This approach may
help
to characterize additional odd transport phenomena with divergence-free fluxes, such as odd heat conduction and odd couplings between viscous and diffusive transport.

\vspace{0.1in}
\begin{acknowledgments}
\noindent\textbf{\textit{Acknowledgements.}}
C.H. is supported by the National Science Foundation, Division of Chemical, Bioengineering, Environmental, and Transport Systems (CBET) under award number $2039624$, and by the National Science Foundation Graduate Research Fellowship Program under Grant No.\ DGE 1752814. K.K.M is supported by Director, Office of Science, Office of Basic Energy Sciences, of the U.S. Department of Energy under contract No. DEAC02-05CH11231.
\end{acknowledgments}

\bibliography{ref}

\onecolumngrid
\renewcommand{\theequation}{A.\arabic{equation}}
\renewcommand{\thefigure}{A.\arabic{figure}}
\renewcommand{\thesection}{A.\arabic{section}}

\newpage
\begin{center}
\textbf{Appendices} \\
\vspace{0.05in}
\end{center}

\setcounter{equation}{0}
\setcounter{footnote}{0}
\setcounter{section}{0}
\setcounter{figure}{0}

\section{Chiral random walk}\label{section:appendix-chiral-random-walk}
In this appendix, we present derivations of the analytical expressions in the main text concerning the chiral random walk model.
We begin by considering the balance equations for the joint probability densities of the particle occupying coordinates $(x,y)$ at time $t$ while moving in one of the four available directions indicated by $\{\rightarrow, \uparrow, \leftarrow, \downarrow\}$ with fixed speed $v_0$.
For instance,
\begin{align}
\begin{split}
    P_\rightarrow(x,y,t+\delta t) = P_\rightarrow(x - \delta x, y, t) + \delta t \big[&\Gamma_1 P_\downarrow(x,y,t) + \Gamma_2 P_\leftarrow(x,y,t) \\
    + &\Gamma_3 P_\uparrow(x,y,t) - \gamma P_\rightarrow(x,y,t) \big]\,,
\end{split}
\end{align}
where $\delta x = v_0 \delta t$ and $\gamma = \Gamma_1 + \Gamma_2 + \Gamma_3$ is the total turning frequency. Taking the limit $\delta t \rightarrow 0$ and repeating the process for the other directions yields the coupled master equations~\eqref{eq:master-1}-\eqref{eq:master-4}.
We can solve the master equations by applying Fourier and Laplace transforms in space and time, respectively:
\begin{align}
    \label{eq:fourier-laplace-1}
    &(s + \gamma) \tilde{P}_\rightarrow(\bm{k}, s) + \mathrm{i} k_x v_0 \tilde{P}_\rightarrow(\bm{k}, s)
    - \Gamma_1 \tilde{P}_\downarrow(\bm{k}, s)
    - \Gamma_2 \tilde{P}_\leftarrow(\bm{k}, s)
    - \Gamma_3 \tilde{P}_\uparrow(\bm{k}, s) = P_\rightarrow(\bm{k}, 0)\,, \\
    \label{eq:fourier-laplace-2}
    &(s + \gamma) \tilde{P}_\uparrow(\bm{k}, s) + \mathrm{i} k_y v_0 \tilde{P}_\uparrow(\bm{k}, s)
    - \Gamma_1 \tilde{P}_\rightarrow(\bm{k}, s)
    - \Gamma_2 \tilde{P}_\downarrow(\bm{k}, s)
    - \Gamma_3 \tilde{P}_\leftarrow(\bm{k}, s) = P_\uparrow(\bm{k}, 0)\,, \\
    \label{eq:fourier-laplace-3}
    &(s + \gamma) \tilde{P}_\leftarrow(\bm{k}, s) - \mathrm{i} k_x v_0 \tilde{P}_\leftarrow(\bm{k}, s)
    - \Gamma_1 \tilde{P}_\uparrow(\bm{k}, s)
    - \Gamma_2 \tilde{P}_\rightarrow(\bm{k}, s)
    - \Gamma_3 \tilde{P}_\downarrow(\bm{k}, s) = P_\leftarrow(\bm{k}, 0)\,, \\
    \label{eq:fourier-laplace-4}
    &(s + \gamma) \tilde{P}_\downarrow(\bm{k}, s) - \mathrm{i} k_y v_0 \tilde{P}_\downarrow(\bm{k}, s)
    - \Gamma_1 \tilde{P}_\leftarrow(\bm{k}, s)
    - \Gamma_2 \tilde{P}_\uparrow(\bm{k}, s)
    - \Gamma_3 \tilde{P}_\rightarrow(\bm{k}, s) = P_\downarrow(\bm{k}, 0)\,,
\end{align}
where the transforms are defined as
\begin{align}
    \label{eq:fourier-transform}
    f(\bm{x},t) &= \frac{1}{2\pi} \int_{-\infty}^\infty d\bm{k}\ f(\bm{k}, t) e^{\mathrm{i} \bm{k} \cdot \bm{x}}\,, \\
    \label{eq:laplace-transform}
    f(\bm{x},t) &= \int_0^\infty ds\ \tilde{f}(\bm{x}, s) e^{s t} \,.
\end{align}

To quantify $D_\parallel$, we ask how the total probability density $\tilde{P}(\bm{k}, s) = \tilde{P}_\rightarrow(\bm{k}, s) + \tilde{P}_\uparrow(\bm{k}, s) + \tilde{P}_\leftarrow(\bm{k}, s) + \tilde{P}_\downarrow(\bm{k}, s)$ spreads out in time from a point, allowing us to calculate the mean-squared displacement.
To this end, we specify the isotropic initial conditions $P_\rightarrow(\bm{k}, 0) = P_\uparrow(\bm{k}, 0) = P_\leftarrow(\bm{k}, 0) = P_\downarrow(\bm{k}, 0) = 1/4$ and consequently are free to choose any direction for $\bm{k}$. Arbitrarily setting $\bm{k} = k_x\bm{\hat{e}_x}$ and solving algebraically yields
\begin{equation} \label{eq:P-full-solution}
    \tilde{P}(k_x, s) =
    \frac{2 (2 \gamma -2 \Gamma _2+s) \big[(\Gamma _1-\Gamma _3)^2+(\gamma +\Gamma _2+s)^2\big]
    + k_x^2 v_0^2 (\gamma +\Gamma _2+s)}
    {2s (2 \gamma -2 \Gamma _2+s) \big[(\Gamma _1-\Gamma _3)^2+(\gamma +\Gamma _2+s)^2\big]
    +2 k_x^2 v_0^2 \big[(\gamma +s)^2-\Gamma _2^2\big] }\,.
\end{equation}
We may then obtain the second moment of the probability density as
\begin{equation}
    \langle \Delta \tilde{x}(s)^2 \rangle = -\partial_{k_x}^2 \tilde{P}(k_x, s)\big\rvert_{k_x=0}
    =\frac{v_0^2 (\gamma +\Gamma _2+s)}
    {s^2\big[(\Gamma _1-\Gamma _3)^2+(\gamma +\Gamma _2+s)^2\big]}\,.
\end{equation}
Taking the diffusive limit $s \rightarrow 0$ and performing the inverse Laplace transform~\eqref{eq:laplace-transform} yields an expression for the diffusion coefficient $D_\parallel$ from the mean-squared displacement relation in the third equality of~\eqref{eq:gk-1} in the main text:
\begin{equation} \label{eq:D-parallel-msd}
    \lim_{t \rightarrow \infty} \langle \Delta x(t)^2 \rangle = 2 D_\parallel t
    = \bigg(\frac{v_0^2 (\gamma +\Gamma _2)}{(\Gamma _1-\Gamma _3)^2 + (\gamma +\Gamma _2)^2} \bigg)t \,.
\end{equation}

As noted in the main text, because the diffusion equation~\eqref{eq:diffusion-equation} does not involve $D_\perp$, the second moment of $P(x,y,t)$ does not contain any direct information about $D_\perp$.
Instead, from the expansion described in~\eqref{eq:chiral-rw-gk-1}-\eqref{eq:chiral-rw-gk-2}, we may consider the first moment when specifying both the initial position and initial velocity in equations~\eqref{eq:fourier-laplace-1}-\eqref{eq:fourier-laplace-4}.
For example, to obtain $\langle x(t) \rangle_\rightarrow$ we set $P_\rightarrow(\bm{k}, 0) = 1$ and $P_\uparrow(\bm{k}, 0) = P_\leftarrow(\bm{k}, 0) = P_\downarrow(\bm{k}, 0) = 0$, and choose $\bm{k} = k_x\bm{\hat{e}_x}$. Solving equations~\eqref{eq:fourier-laplace-1}-\eqref{eq:fourier-laplace-4} as before and adding to obtain the total probability density yields
\begin{equation}
    \tilde{P}(k_x, s) = \frac{(2\gamma -2\Gamma _2+s) \big[-\mathrm{i} k_x v_0 (\gamma +\Gamma _2+s) + (\Gamma _1-\Gamma _3)^2 +(\gamma +\Gamma _2+s)^2\big]}
    {s (2 \gamma -2 \Gamma _2+s) \big[(\Gamma _1-\Gamma _3)^2+(\gamma +\Gamma _2+s)^2\big] + k_x^2 v_0^2 \big[(\gamma +s)^2-\Gamma _2^2\big]} \,.
\end{equation}
Note that, unlike in equation~\eqref{eq:P-full-solution}, $\tilde{P}(k_x, s)$ now has an imaginary part due to the asymmetry of the initial conditions. Differentiating in $k_x$ obtains the first moment
\begin{equation}\label{eq:x-moment}
    \langle \tilde{x}(s) \rangle_{\rightarrow} = \mathrm{i} \partial_{k_x} \tilde{P}(k_x, s)\big\rvert_{k_x=0}
    = \frac{v_0 (s + \gamma + \Gamma _2)}
    {s\big[(s + \gamma + \Gamma_2)^2 + (\Gamma_1 - \Gamma_3)^2 \big]}\,.
\end{equation}
Taking the same approach but choosing instead $\bm{k} = k_y\bm{\hat{e}_y}$, we find
\begin{equation}\label{eq:y-moment}
    \langle \tilde{y}(s) \rangle_{\rightarrow} = \mathrm{i} \partial_{k_y} \tilde{P}(k_y, s)\big\rvert_{k_y=0}
    = \frac{v_0 (\Gamma_1 - \Gamma_3)}
    {s\big[(s + \gamma + \Gamma_2)^2 + (\Gamma_1 - \Gamma_3)^2 \big]}\,.
\end{equation}
Finally, introducing the notation $\omega = \Gamma_1 - \Gamma_3$ and $\nu = \Gamma_1 + 2\Gamma_2 + \Gamma_3$, and performing the inverse Laplace transform~\eqref{eq:laplace-transform} on equations~\eqref{eq:x-moment}-\eqref{eq:y-moment} leads to the logarithmic spiral form given in~\eqref{eq:CRW-x-average}-\eqref{eq:CRW-y-average}.
The diffusion coefficients $D_\parallel$ and $D_\perp$ given in equations~\eqref{eq:Dpar-crw}-\eqref{eq:Dperp-crw} then follow directly from the long-time response as $t \rightarrow \infty$.

\begin{figure}[!b]
    \centering
    \includegraphics[width=.7\textwidth]{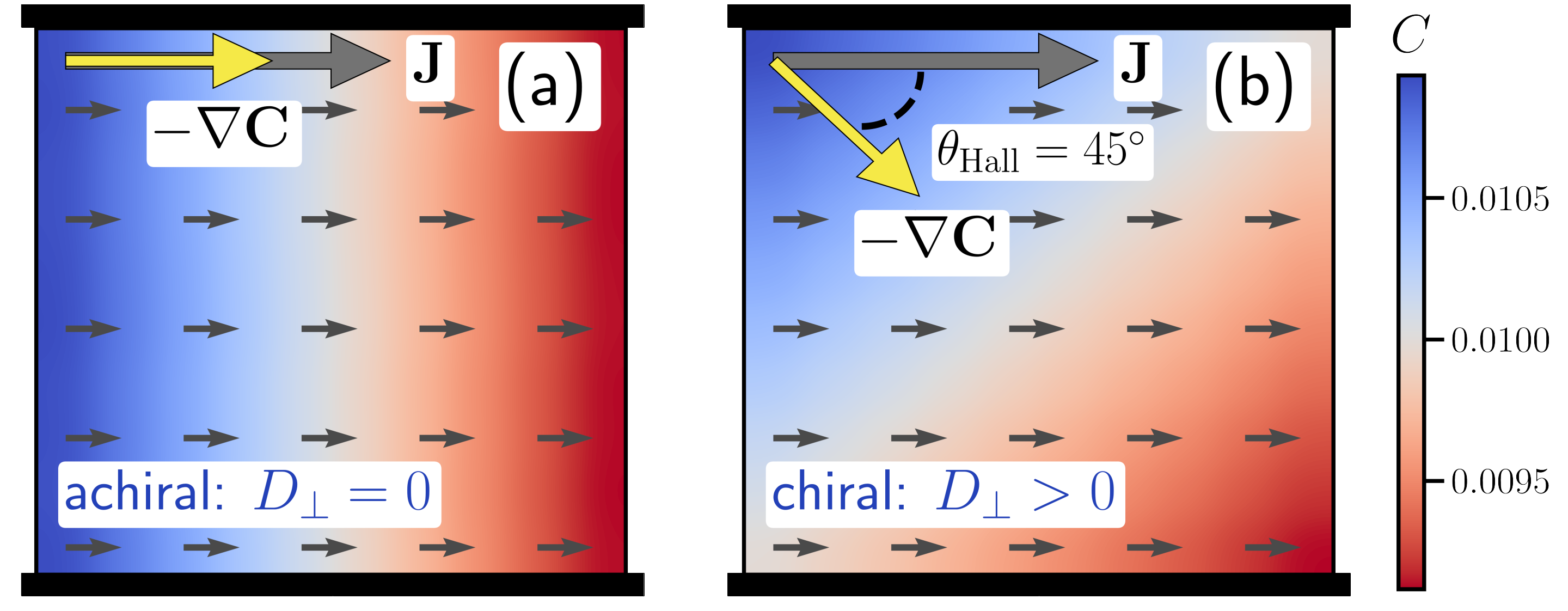}
    \caption{
    Steady-state concentration profile for diffusive flux through a channel with impermeable walls obtained from numerical simulation of the chiral random walk model without odd diffusivity (a; achiral, $\Gamma_1=1, \Gamma_2=0, \Gamma_3=1$) and with odd diffusivity (b; chiral, $\Gamma_1=1, \Gamma_2=0, \Gamma_3=0$).
    }
    \label{fig:channel-flux-boundaries}
\end{figure}
\vspace{5pt}
One can understand the effect odd diffusivity may have on the concentration by constructing a boundary value problem.
Let us consider a channel of length $L$ whose top and bottom boundaries are impermeable and separated by a distance $W$, and to which particles are added at the left boundary and removed from the right boundary at a constant rate $J_0 W$.
These boundary conditions suggest the ansatz $\bm{J}(x,y) = J_0 \hat{\bm{e}}_x$ for all $(x,y)$.
Then, from the constitutive relations of~\eqref{eq:ficks-law} and~\eqref{eq:diffusion-tensor}, we have
\begin{align}
        \label{eq:channel-flux-ansatz-1}
        J_0 &= -D_\parallel \partial_x C + D_\perp \partial_y C\,, \\
        \label{eq:channel-flux-ansatz-2}
        0 &= -D_\parallel \partial_y C - D_\perp \partial_x C \,.
\end{align}
Upon defining the average concentration $C_0 = \frac{1}{LW}\int_0^L dx \int_0^W dy\ C(x,y) = C(0,0)$, equations~\eqref{eq:channel-flux-ansatz-1}-\eqref{eq:channel-flux-ansatz-2} permit the solution
\begin{align} \label{eq:channel-flux}
    \begin{split}
        C^\mathrm{ss}(x,y) &= C_0 + \frac{J_0}{D_\parallel^2 + D_\perp^2} \big( -D_\parallel x + D_\perp y \big) \\
        &= C_0 + \frac{J_0}{v_0^2} \big( -\nu x + \omega y \big) \,.
    \end{split}
\end{align}
When $D_\perp \ne 0$, as seen from equation~\eqref{eq:channel-flux}, asymmetric accumulation occurs along the impermeable channel walls giving rise to a linear concentration profile not only in the $x$-direction but in the $y$-direction as well.

We check this solution by running numerical simulations of the chiral random walk model with corresponding boundary conditions, where the probability density $P$ is interpreted as the concentration $C$.
Specifically, we simulate the dynamics of a particle governed by equations~\eqref{eq:master-1}-\eqref{eq:master-4} with either $\Gamma_1 = 1, \Gamma_2=0, \Gamma_3=1$ (left- and right-turning) or $\Gamma_1 = 1, \Gamma_2=0, \Gamma_3=0$ (left-turning only) for a single particle in a box of dimensions $L=10$, $W=10$, advancing the dynamics in timesteps of $\delta t = 0.01$.
Whenever the particle crosses the boundary at $x = L$, it is replaced at $x=0$ on the next timestep.
In Figure~\ref{fig:channel-flux-boundaries} we plot the steady-state simulation average, finding the resulting flux field to be uniform while the concentration field depends linearly on $x$ and $y$, in agreement with equation~\eqref{eq:channel-flux} where $C_0 = 0.01$ and $J_0 = 0.0001$.

\section{Molecular dynamics simulation details} \label{section:appendix-simulation-details}
Molecular dynamics simulations of a passive tracer particle diffusing in a chiral active bath composed of self-spinning dumbbells were performed in LAMMPS~\cite{lammps} with custom modifications\footnote{Our simulation and analysis code is publicly available at https://github.com/mandadapu-group/active-matter.} implementing the microscopic active forces and constant-flux boundary conditions.
The nonconservative active force $\bm{f}^A_i$ in equation~\eqref{eq:langevin-dynamics} affects only the dumbbell particles, with constant magnitude $|\bm{f}^A_i| = f^A$.
The orientation of $\bm{f}^A_i$ is perpendicular to the bond vector $\bm{r}_i - \bm{r}_j$ for the bonded pair $i$ and $j$, and directed oppositely ($\bm{f}^A_i = -\bm{f}^A_j$), inducing rotation of the dumbbell.
Chiral active dumbbells are composed of two particles held together by a harmonic potential $U^\mathrm{Harm}(r) = \frac{1}{2}k(r-r_0)^2$, where $r$ is the separation distance.
We set the spring constant $k=100$ and the reference bond length $r_0 = 1$.
All particles (including the passive tracer) interact with non-bonded neighbors through a Weeks-Chandler-Andersen~\cite{Weeks71} potential defined by
\begin{equation}
U^\mathrm{WCA}(r) = \begin{cases}
      4\epsilon \bigg[ \big(\sigma/r\big)^{12} - \big(\sigma/r\big)^6 \bigg] + \epsilon & r < 2^{1/6}\sigma \\
      0 & r \geq 2^{1/6}\sigma \,, \\
   \end{cases}
\end{equation}
such that $U = U^\mathrm{Harm} + U^\mathrm{WCA}$ in equation~\eqref{eq:langevin-dynamics}.
Here, $m$, $\sigma$ and $\epsilon$ are the particle mass, diameter and interaction energy, providing characteristic mass, length and energy scales which define the Lennard-Jones units system.
All simulation results are reported in Lennard-Jones units.
The Langevin dynamics described in equation~\eqref{eq:langevin-dynamics} were discretized with a velocity Verlet scheme with time step $\delta t = 0.005$ and bath temperature $k_\mathrm{B} T = 1.0$.
The friction coefficient was set to $\zeta = 2.0$ for dumbbell particles and $\zeta = 0$ for the passive tracer particles, such that the tracers move ballistically between collisions.
Simulations were performed at high dilution of the passive solute particles, where all simulations contained at least twenty times the number of active dumbbell solvent particles as passive tracer solute particles.

\begin{figure}[!t]
    \centering
    \includegraphics[width=.8\textwidth]{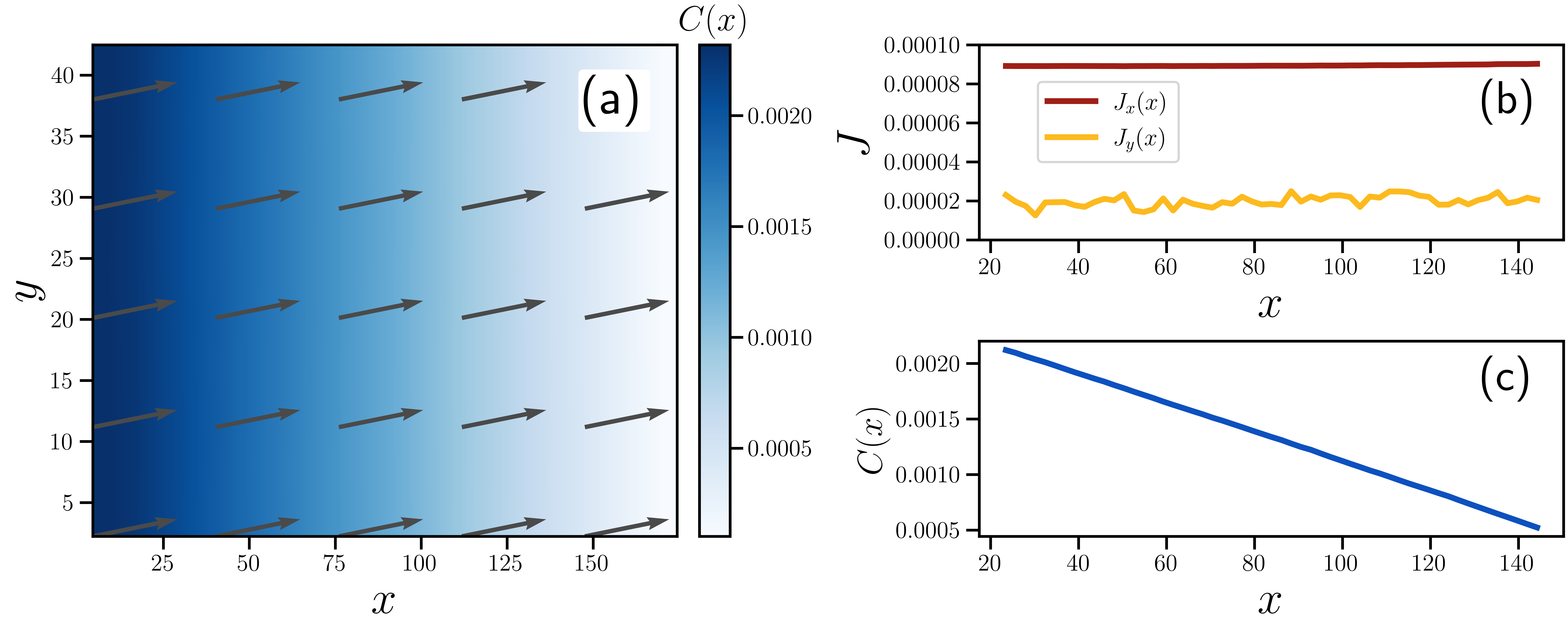}
    \caption{Results of a typical boundary-driven flux simulation of diffusion of a passive tracer particle in a chiral active bath. Parameters $\rho_\mathrm{active}=0.1$ and $\mathrm{Pe} = 16$ have been chosen arbitrarily. (a) The flux field (arrows) is spatially homogeneous with a component in the $y$-direction due to odd diffusivity, while the concentration $C(x)$ varies linearly in the $x$-direction. The profiles of the flux and the concentration along the $x$-direction are plotted in (b) and (c), respectively. All quantities are averaged over $2 \times 10^8$ timesteps.
    }
    \label{fig:boundary-driven-flux}
\end{figure}

Calculation of the velocity autocorrelation tensor entering the Green-Kubo relations~\eqref{eq:gk-1} and~\eqref{eq:gk-2} was performed in a fully periodic system in a non-equilibrium steady state exhibiting stationarity and spatial homogeneity of all observables. Boundary-driven flux simulations were performed in a rectangular simulation box with special boundary conditions affecting the diffusing passive solute particles but not the active bath particles. A passive solute particle passing out of the simulation box through the right boundary behaves periodically, reappearing at the left boundary. A passive solute particle particle passing through the left boundary is reflected back into the simulation box. All interactions across the boundaries remain fully periodic. These conditions ensure a constant flux of particles across the simulation box, with the concentration varying linearly in $x$, as shown in Figure~\ref{fig:boundary-driven-flux} for a particular simulation with $\rho_\mathrm{active}=0.1$ and $\mathrm{Pe} = 16$.

\section{Linear response mobility tensor}\label{section:appendix-mobility}
\begin{figure}[!t]
    \centering
    \includegraphics[width=.8\textwidth]{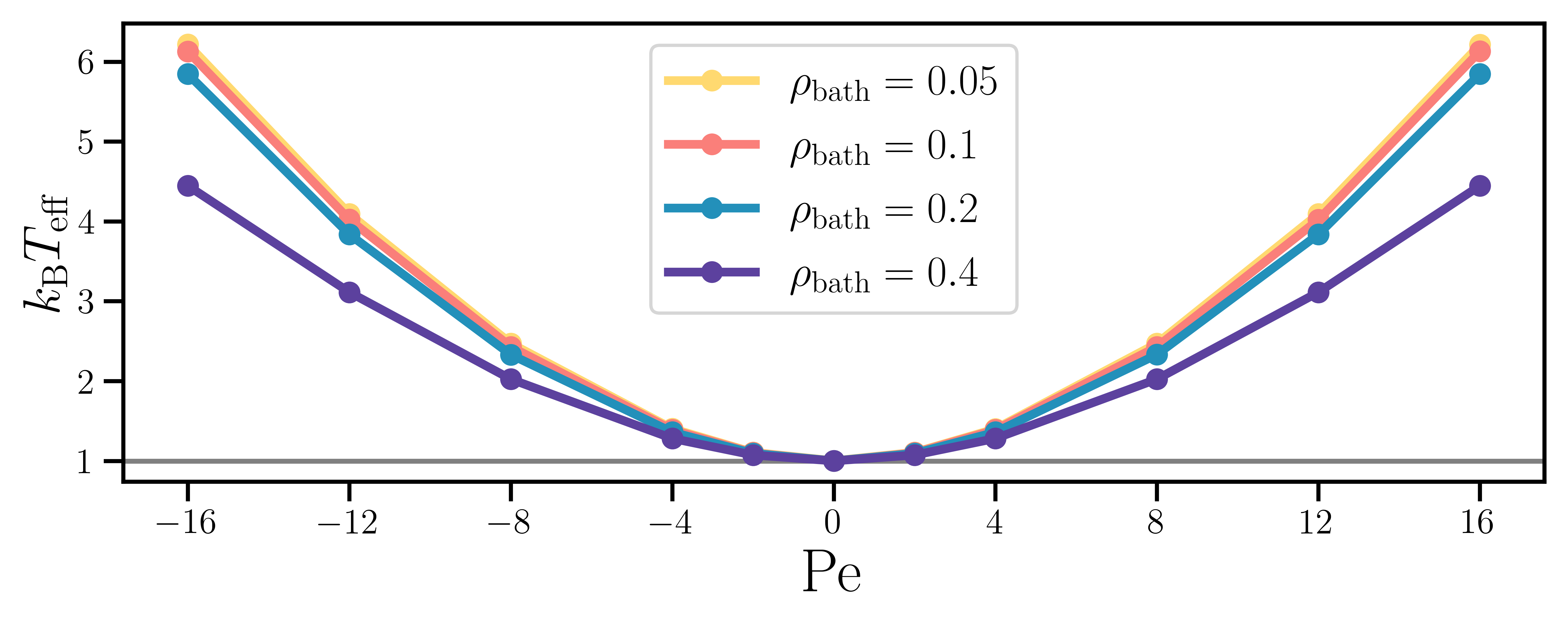}
    \caption{
    Effective kinetic temperature of the passive tracer particle across all values of $\rho_\mathrm{bath}$ and $\mathrm{Pe}$ corresponding to the simulation results displayed in Figure~\ref{fig:gk_vs_nemd} of the main text.
    }
    \label{fig:effective_temperature}
\end{figure}
The mobility tensor $\boldsymbol{\mu}$ provides a linear relation between a particle's drift velocity $\bm{u}$ and an applied body force $\bm{g}$ which, within the context of linear response theory, is expected to be valid for sufficiently small $\bm{g}$
\begin{equation}
    u_i = \mu_{ij} g_j \,.
\end{equation}
For passive systems, the mobility and diffusivity are ordinarily connected by the Einstein relation
\begin{equation} \label{eq:einstein-relation}
    D_{ij} = k_\mathrm{B} T \mu_{ij} \,.
\end{equation}
Active matter systems need not obey such a relation.
Indeed, one of the hallmarks of many active matter models is the ``enhancement'' of the diffusivity, due to the presence of active driving forces, over its value in the absence of such forces.
When such behavior is present, the Green-Kubo relations for the diffusivity coefficients in equations~\eqref{eq:gk-1}-\eqref{eq:gk-2} are expected to remain valid while predictions of the diffusivity coefficients from linear response theory \textit{via} the Einstein relation~\eqref{eq:einstein-relation} cease to be applicable.

\begin{figure}[!t]
    \centering
    \includegraphics[width=.48\textwidth]{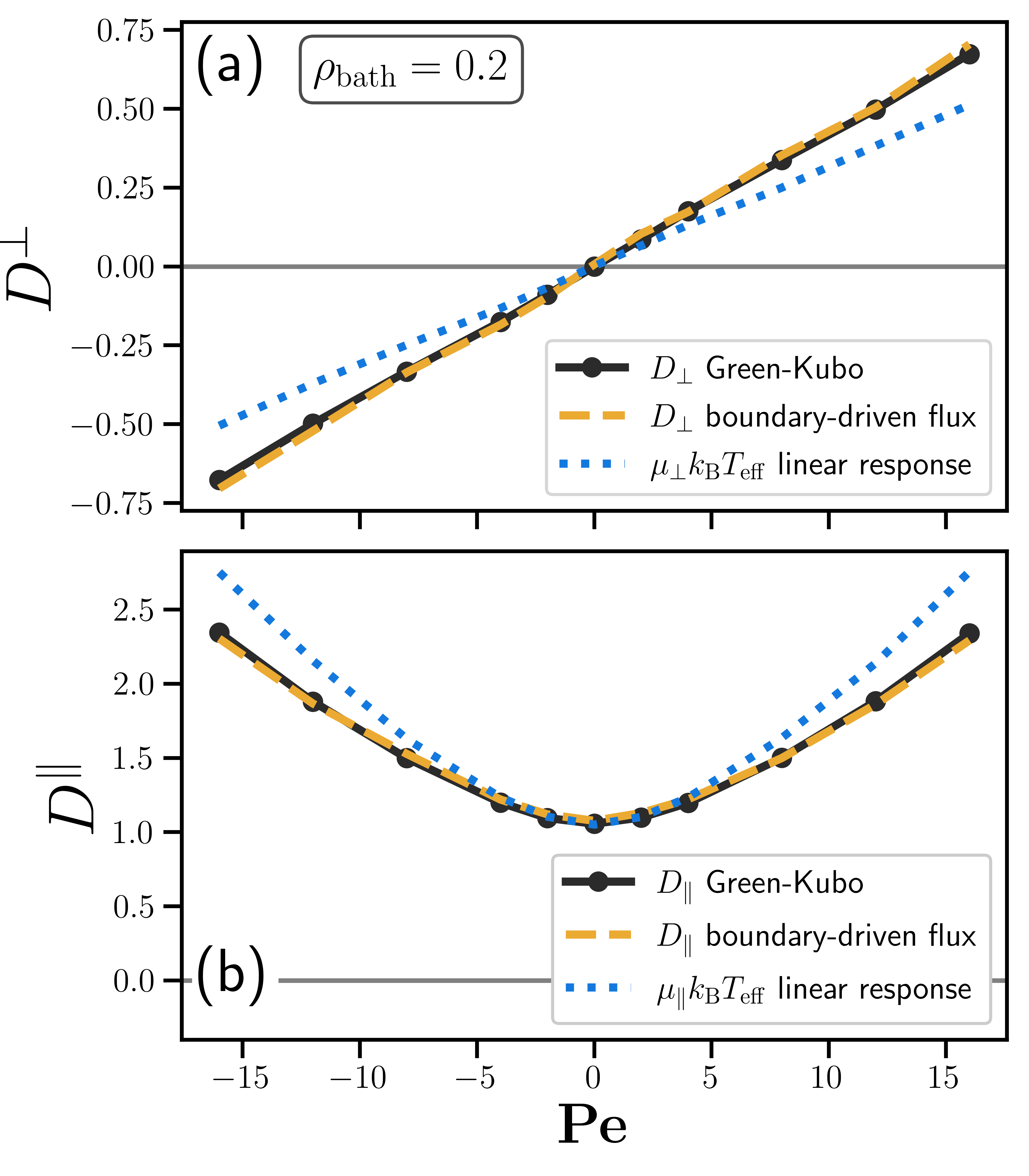}
    \caption{Comparison of the diffusion coefficients for a passive tracer particle in an active dumbbell bath with $\rho_\mathrm{bath} = 0.2$ obtained from Green-Kubo and boundary-driven flux calculations (solid lines and dashed lines, respectively) against those predicted from the from the mobility using the Einstein relation with an effective kinetic temperature.
    }
    \label{fig:gk_vs_nemd_vs_body}
\end{figure}

To illustrate this, let us briefly consider a simple model system which exhibits nonzero odd diffusivity but whose mobility tensor contains no antisymmetric part.
Namely, we consider an active Brownian particle in two dimensions in the overdamped regime, driven by internally-generated forces oriented along a director $\hat{\bm{u}}(t) = \big(\cos \theta(t),\ \sin \theta(t) \big)$, where $\theta(t)$ is the polar angle of the director.
We consider the case where the evolution of the director has both a random part, due to interactions with the environment or internal noise, as well as a deterministic bias, due to an internally generated torque.
This setup has been suggested as a minimal model for zooplankton such as \textit{Daphnia}, which tend to steer either left or right as they swim in-plane~\cite{romanczuk2012active,Komin2004}.
The Langevin equations for such a system are
\begin{align}
    \label{eq:langevin-cABP-1}
    \dot{\bm{r}} &= v_0 \hat{\bm{u}}\,, \\
    \label{eq:langevin-cABP-2}
    \dot{\theta} &= \omega_0 + \sqrt{2 D_r} \xi_r(t)\,,
\end{align}
where $\xi_r(t)$ is Gaussian white noise characterized by $\langle \xi_r(t) \rangle = 0$ and $\langle \xi_r(t) \xi_r(t') \rangle = \delta(t-t')$.

The velocity correlation functions for this isotropic system are
\begin{align}
    \langle v_x(t) v_x(0) \rangle = \langle v_y(t) v_y(0) \rangle &= v_0^2 \langle \cos \theta(t) \cos \theta(0) \rangle \\
    \langle v_x(t) v_y(0) \rangle = -\langle v_y(t) v_x(0) \rangle &= v_0^2 \langle \cos \theta(t) \sin \theta(0) \rangle
\end{align}
Using trigonometric product identities, one may show that
\begin{align}
    \label{eq:cABP-correlator-1}
    \langle \cos \theta(t) \cos \theta(0) \rangle &= \frac{1}{2} \langle \cos \big(\theta(t) - \theta(0)\big) + \cos \big(\theta(t) + \theta(0)\big) \rangle
    = \frac{1}{2} \langle \cos \phi(t) \rangle \\
    \label{eq:cABP-correlator-2}
    \langle \cos \theta(t) \sin \theta(0) \rangle &= \frac{1}{2} \langle \sin \big(\theta(t) + \theta(0)\big) - \sin \big(\theta(t) - \theta(0)\big) \rangle
    = -\frac{1}{2} \langle \sin \phi(t) \rangle
\end{align}
where the second equality in both equations follows from isotropy and $\phi(t) = \theta(t) - \theta(0)$ is the displacement at time $t$ of the angle from its initial value.

The Fokker-Planck equation  corresponding to the Langevin equation~\eqref{eq:langevin-cABP-2} is~\cite{Risken1989}
\begin{equation}\label{eq:cABP-fokker-planck}
    \frac{\partial}{\partial t} f(\phi, t) = \omega_0 \frac{\partial}{\partial \phi} f(\phi, t) + D_r \frac{\partial^2}{\partial \phi^2} f(\phi, t)\,,
\end{equation}
where $f(\phi, t)$ is the probability density of the director angle.
Defining the characteristic function of the angle distribution as
\begin{equation}
    \tilde{f}(k, t) = \langle e^{\mathrm{i} k \phi}\rangle = \int_{-\infty}^{\infty} d\phi\ e^{\mathrm{i} k \phi} f(\phi, t)\,,
\end{equation}
Equation~\eqref{eq:cABP-fokker-planck} can be solved in Fourier space resulting in
\begin{equation}
    \tilde{f}(k, t) = \exp\big[ (\mathrm{i} k \omega_0 - k^2 D_r ) t \big] \,.
\end{equation}
Thus,
\begin{align}
    \label{eq:cABP-conditional-1}
    \langle \cos \phi(t) \rangle &= \operatorname{Re} \tilde{f}(1, t) = \cos ( \omega_0 t) e^{-D_r t} \,, \\
    \label{eq:cABP-conditional-2}
    \langle \sin \phi(t) \rangle &= \operatorname{Im} \tilde{f}(1, t) = \sin ( \omega_0 t) e^{-D_r t}
\end{align}
Finally, inserting equations~\eqref{eq:cABP-correlator-1}-\eqref{eq:cABP-correlator-2} and \eqref{eq:cABP-conditional-1}-\eqref{eq:cABP-conditional-2} into the Green-Kubo relations~\eqref{eq:gk-1}-\eqref{eq:gk-2} yields
\begin{align}
    D_\parallel = \frac{v_0^2}{2} \frac{D_r}{D_r^2 + \omega_0^2}\,, \\
    D_\perp = \frac{v_0^2}{2} \frac{\omega_0}{D_r^2 + \omega_0^2}\,.
\end{align}
Note that the functional form is identical to that of the chiral random walk model in equations~\eqref{eq:Dpar-crw}-\eqref{eq:Dperp-crw}, elucidating the merits of this model in capturing the essential features of the odd diffusivity.
Now, as the mechanisms generating active propulsive forces and steering torques were assumed to be ``internal'', \textit{i.e.} not resulting from external interactions, the mobility tensor in this idealized model will be symmetric and independent of the values of $v_0$ and $\omega_0$, for instance following Stokes' Law.

We now evaluate the applicability of an effective Einstein relation for the chiral active dumbbell bath model discussed in the main text, upon defining an effective temperature computed from the mean kinetic energy of the diffusing passive tracer particle~\eqref{eq:einstein-relation}:
\begin{equation}
    k_\mathrm{B} T_\mathrm{eff} = \frac{1}{2} \langle |\bm{v}_\mathrm{tracer}|^2 \rangle\,.
\end{equation}
The dependence of this temperature on $\mathrm{Pe}$ is plotted for all densities of the dumbbell bath in Figure~\ref{fig:effective_temperature}, corresponding to the simulation results plotted in Figure~\ref{fig:gk_vs_nemd} of the main text.
The temperature of the nonequilibrium stationary state is determined by the competition between active forces and dissipative Langevin forces and, more noticeably at higher dumbbell densities, collisions occurring between dumbbells.

The resulting relationship is plotted in Figure~\ref{fig:gk_vs_nemd_vs_body}, where we have defined the isotropic mobility tensor analogously to the diffusivity as $\mu_{ij} = \mu_\parallel \delta_{ij} - \mu_\perp \epsilon_{ij}$. We observe that the linear response prediction captures only the qualitative behavior of $D_\perp$ and $D_\parallel$, with the disagreement most pronounced at high $\mathrm{Pe}$.
Note, finally, that because the sign of the linear response error differs for $D_\perp$ and $D_\parallel$ in Figure~\ref{fig:gk_vs_nemd_vs_body}, no single choice of $T_\mathrm{eff}$ could simultaneously reconcile the disagreement for both diffusion coefficients.

\end{document}